\documentclass[reviewcopy]{elsart}
\usepackage{amssymb}
\usepackage{amsmath}
\usepackage{graphicx}

\newcommand{\bmath}[1]{\mbox{\boldmath ${#1}$}}

\newcommand{\AAA}{\textrm{A}}
\newcommand{\AC}{\textrm{C}}

\newcommand{\AAu}{\textrm{Au}}

\begin{document}

\begin{frontmatter}

\vspace*{-10mm}
\title{\large{\bf Evidence of kaon nuclear and  Coulomb potential
effects on soft $\bmath{K^+}$ production from nuclei.}}

\author[juelich,gatchina]{M.~Ne\-ki\-pe\-lov\thanksref{CA}},
\author[juelich]{M.~B\"uscher},
\author[giessen]{W.~Cassing},
\author[juelich]{M.~Hartmann},
\author[juelich]{V.~Hejny},
\author[juelich]{V.~Kleber},
\author[juelich]{H.R.~Koch},
\author[gatchina]{V.~Koptev},
\author[juelich]{Y.~Maeda},
\author[juelich]{R.~Maier},
\author[dubna]{S.~Merzliakov},
\author[gatchina]{S.~Mi\-kir\-tychiants},
\author[juelich]{H.~Ohm},
\author[dubna]{A.~Petrus},
\author[juelich]{D.~Prasuhn},
\author[juelich]{F.~Rathmann},
\author[juelich,cracow]{Z.~Rudy},
\author[juelich]{R.~Schleichert},
\author[juelich]{H.~Schneider},
\author[juelich]{K.~Sistemich},
\author[juelich]{H.J.~Stein},
\author[juelich]{H.~Str\"oher},
\author[juelich]{K.-H.~Watzlawik},
\author[london]{C.~Wilkin}

\thanks[CA]{Corresponding author. E-mail address: m.nekipelov@fz-juelich.de}

\address[juelich]{Institut f\"ur Kernphysik, Forschungszentrum J\"ulich,
                  D-52425 J\"ulich, Germany}
\address[gatchina]{High Energy Physics Department, Petersburg Nuclear Physics
                    Institute, 188350 Gatchina, Russia}
\address[giessen]{Institut f\"ur Theoretische Physik, Justus Liebig 
                  Universit\"at Giessen, D-35392 Giessen, Germany}
\address[dubna]{Laboratory of Nuclear Problems, Joint Institute for Nuclear
         Research, Dubna, 141980 Dubna, Moscow Region, Russia}
\address[cracow]{M. Smoluchowski Institute of Physics, Jagellonian
                  University, Reymonta 4, PL-30059 Cracow, Poland}
\address[london]{Physics Department, UCL, Gower Street,
                    London WC1 6BT, England}

\begin{abstract}
  The ratio of forward $K^+$ production on copper, silver and gold
  targets to that on carbon has been measured at proton beam energies
  between 1.5 and 2.3 GeV as a function of the kaon momentum $p_K$
  using the ANKE spectrometer at COSY-J\"ulich. The strong suppression
  in the ratios observed for $p_K<200$--$250$~MeV/c may be ascribed to
  a combination of Coulomb and nuclear repulsion in the $K^+A$ system.
  This opens a new way to investigate the interaction of $K^+$-mesons
  in the nuclear medium. Our data are consistent with a $K^+A$ nuclear
  potential of $V_K^0\approx 20\,$MeV at low kaon momenta and normal
  nuclear density. Given the sensitivity of the data to the kaon
  potential, the current experimental precision might allow one to
  determine $V_K^0$ to better than 3$\,$MeV.
\end{abstract}

\end{frontmatter}

Final state interactions of $K^+$ mesons in nuclei are generally
considered to be rather small, due to their strangeness of $S= +1$.
As a consequence, the production of $K^+$-mesons in proton-nucleus
collisions is of great importance to learn about either
cooperative nuclear phenomena or high momentum components in the
nuclear many-body wave function. This is particularly the case since
the production of kaons, being relatively heavy as compared to pions,
requires strong medium effects.

Several groups have made experimental and theoretical studies of total
and doubly-differential $K^+$-production cross sections over a wide
range of proton beam
energies~\cite{Ko88,Schn88,Ca90,Ca94,Si94,MB96,De96,Ba98,Ki99,Pa00,Ko01,MB02}.
These studies show that a two-step reaction mechanism, involving the
production of an intermediate $\Delta$ or $\pi$, dominates below the
threshold of the elementary $pN\to K^+\Lambda N$ reaction
($T_p=1.58$~GeV).  A strong target mass dependence of the production
rate may be a good indicator for the dominance of such secondary
mechanisms. At higher energies the role of the secondary effects
decreases, especially in the high momentum part of the kaon spectra,
where direct production dominates~\cite{Rudy}.

It is, however, clear that the repulsive Coulomb potential in the
target nucleus will distort the soft part of the momentum spectrum.
Furthermore, since the $K^+$ nuclear potential, though small, is also
repulsive~\cite{Si97}, with a strength rather similar to that of the
Coulomb for a heavy nucleus, this distortion will be reinforced. For
this reason there have been several publications which have stressed
the importance of including the effects of Coulomb and nuclear
potentials on the propagation of mesons in the nuclear
medium~\cite{Si97,Ay97,Te97}. Such effects can change the
interpretation of the shape of the $K^+$ spectrum as well as of the
mass dependence of the cross sections; therefore they have to be taken
into account in the interpretation of the experimental results.

The influence of final state rescattering effects in meson production
can be investigated experimentally for high momentum mesons by
measuring two-body reactions, \textit{e.g.\/} $pA\to (A+1)^*\,\pi^+$
for pions or $pA\to\, _{\Lambda}{(A+1)}^*{K^+}$ for kaons, using high
precision spectrometers. However, just as for $\beta^{\pm}$-decay,
much stronger Coulomb effects are expected in the very low momentum
part of the meson spectrum, but there have as yet been no direct
experimental tests, at least for kaons. 
A reliable way of studying this phenomenon is by
measuring directly ratios of cross sections for different nuclei,
since many of the possible systematic errors cancel out.  Measurements
of cross section ratios for mesons of different charges,
\textit{e.g.\/} $\pi^+/\pi^-$ or $K^+/K^-$ are, as a rule, clouded by
the differences in reaction mechanisms. Since most of the measurements
were carried out for high pion and kaon momenta, the existing
experimental data~\cite{C075,Cra80,Ab88,Che02} are not very
informative regarding both Coulomb and kaon potentials. This is
changing with the commissioning of the ANKE spectrometer, which is
currently the only device that is able to measure $K^+$ mesons with
momenta down to $\approx 150$~MeV/c.

Measurements of $K^+$ momentum spectra resulting from proton-nucleus
collisions have been performed with the ANKE spectrometer~\cite{Ba01}
at the COooler SYnchrotron COSY-J\"ulich.  A detailed description of
the kaon detection system is given in Ref.~\cite{Bu01}. The
criteria for the kaon identification and the procedure of measurements
are briefly described as follows: The COSY proton beam, with an
intensity of $(2-4)\times 10^{10}$ protons and a cycle time of $\sim
60\,$s, was accelerated to the desired energy in the range $T_p=1.5 -
2.3$ GeV on an orbit below the target. The targets were thin strips of
C, Cu, Ag or Au with a thickness of (40--1500)$\,\mu$g/cm$^2$.  Over a
period of $\sim 50\,$s, the beam was slowly brought up to the target
by steerers, keeping the trigger rate in the detectors nearly constant
at (1000--1500)$\,$s$^{-1}$, a level that could be handled by the data
acquisition system with a dead time of less than 25\%.  Ejectiles with
horizontal angles in the range $\pm 12^{\circ}$, vertical angles up to
$\pm 7^{\circ}$, and momenta between 150 and 600$\,$MeV/c, were
deflected by the ANKE dipole magnet, passed one of 23 plastic
scintillation counters and 6 planes of $2.5\,$mm wire-step MWPCs. They
were then focussed onto one of the 15 kaon range telescopes, each
consisting of three plastic scintillator counters (stop, $\Delta E$
and veto) and two degraders.  The 10$\,$cm wide telescopes, placed at
the focal surface of the ANKE spectrometer, defined $\sim10$\%
momentum bites in ejectile momenta.  The momentum ranges covered by
each telescope were kept constant for the different beam energies by
operating ANKE at constant magnetic field strength and maintaining the relative
target-dipole-detector geometry.  The thicknesses of the scintillators
and degraders in the telescopes were chosen so as to stop kaons in the
degrader in front of the veto counters. The kaons subsequently decay
with  a mean life time of $\approx 12.4\,$ns and the products of
this decay (pions or muons) are detected by the veto counters with a
delay of more than 1.3$\,$ns with respect to the signals from the
corresponding stop counters.  The combination of the time-of-flight
between the start and stop counters, energy losses in all the
scintillators, delayed particle signals, and information from MWPCs
resulted in clean kaon spectra, with a background of less than 10\% for
all beam energies of 1.5$\,$GeV and above.  The MWPC track
reconstruction allowed us also to vary the angular and momentum
acceptances of the individual telescopes.

The ratio $R(\AAA/\AC)_{p_K}$ of the kaon production cross sections
from heavy (A) to carbon (C) targets for a given kaon momentum $p_K$
can be calculated from the observed number of kaons $n(K^+)$ in the
individual telescopes as:
\begin{equation}
  R(\AAA/\AC)_{p_K}= \left[\frac{n_A(K^+)}{n_C(K^+)}\right]_{p_K}
  \times \frac{L_C}{L_A}\:\cdot
\label{eq:1}
\end{equation}
$L_C$ and $L_A$ denote the integrated luminosities
during data taking with a particular target.  The luminosity ratio
could be obtained from the number $n(\pi^+)$ of 500$\,$MeV/c pions,
measured during pion calibration runs for every energy and for each
target.  
\begin{equation}
\frac{L_C}{L_A}=
\left[\frac{\AAA}{\AC}\right]^{1/3}
\times
\left[\frac{n_C(\pi^+)}{n_A(\pi^+)}\right]_
{p_{\pi}=500\,\textrm{\scriptsize MeV/c}}
\:\cdot
\label{eq:2}
\end{equation}
All numbers of detected pions and kaons in Eq.~(\ref{eq:1}) and
Eq.~(\ref{eq:2}), $n_C$ and $n_A$, were individually normalised to the
relative luminosities during the corresponding runs.  This relative
normalisation was obtained by monitoring the interaction of the proton
beam with the target to an accuracy of 2\% using stop counters 2--5 in
four-fold coincidence directly looking at the target~\cite{Bu01},
thereby selecting ejectiles, produced in the target by hadronic
interactions, which bypassed the spectrometer dipole. Pion production
cross sections in proton-nucleus reactions have been measured by
several groups in the forward direction in the 0.73--4.2$\,$GeV energy
range~\cite{Cra80,Ab88,Co75}.  The combined analysis of these data
showed that, to within 10\%, the ratios of the pion production cross
sections can be scaled with the target mass number as
$\AAA^{1/3}$~\cite{Ba00-Cracow}, which is used in Eq.(\ref{eq:2}).

The absolute values of the doubly-differential cross sections can be
obtained from the numbers of kaons $n(K^+)$ identified by each
telescope, after correction for luminosities, detection efficiencies
in the scintillators and MWPCs, kaon decay between the target and the
telescopes, and angle-momentum acceptances. For the cross section
ratios, used in our present analysis, absolute values are not
needed and many uncertainties of the efficiency corrections cancel out.
The presentation of normalised cross sections is deferred until a later
publication.

\input epsf
\begin{figure}[ht]
  \begin{center}
    \mbox{\epsfxsize=13cm \epsfbox{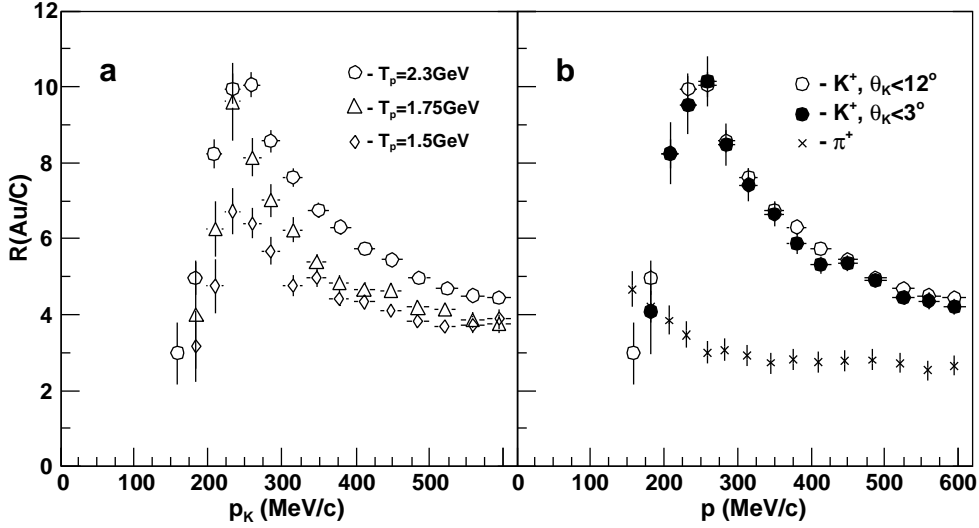}}
    \caption{Ratios of $K^+(\pi^+)$ production cross sections for Au and C
      measured at different beam energies (left figure) and  
      kinematic conditions (right figure) as functions of the
      laboratory meson momentum. a) All the symbols correspond to kaons
      measured in the full ANKE acceptance ($\theta<12^{\circ}$). 
      b) The open circles are as in the left figure, whereas the
      closed circles correspond to kaons measured in the restricted angular interval 
      ($\theta<3^{\circ}$). The $\pi^+$-production cross-section
      ratios of Au to C at $T_p=2.3\,$GeV are designated by crosses.}
  \end{center}
  \label{fig:1}
\end{figure}

The gold/carbon ratio is shown in Fig.~1a for proton beam energies of
2.3, 1.75, and 1.5$\,$GeV. This ratio has a broadly similar shape at
all three energies, with clear maxima for $p_{\textrm{\scriptsize
max}}\approx 245\,$MeV/c coinciding within 2$\,$MeV/c. 
For higher kaon momenta the ratios
decrease monotonically with $p_K$ and in this region the $K^+$
production in gold is relatively stronger at 2.3$\,$GeV than at lower
energies, reflecting changes in the production mechanism with
bombarding energy.  For low kaon momenta one sees a dramatic fall in
the ratio $R(\AAu/\AC)$. To ensure that this phenomenon is not an
artefact of the ANKE detection system, the 2.3$\,$GeV run was repeated
with a reduced dipole magnetic field, resulting in a change in the
values of the momenta that are focussed onto individual range telescopes. The
low $p_K$ suppression remained unchanged. Identical ratios were also
obtained when the polar $K^+$ emission angles in data analysis were 
restricted to lie below $\vartheta_K=3^{\circ}$ with the help of the 
MWPC information (see Fig.1b). 

Ratios of kaon-production cross sections for copper, silver and gold targets 
measured at 2.3$\,$GeV are presented in Fig.~2.
All data exhibit similar shapes, rising steadily with decreasing kaon 
momenta, passing a maximum and falling steeply at low momenta.
The position of the maximum varies with the nucleus, a fit to the data
results in $p_{\textrm{\scriptsize max}}(\AAA/\AC)=245\pm 5$, $232\pm 6$, and
$211\pm 6\,$MeV/c for Au, Ag, and Cu, respectively. The error bars
include contributions from a systematic uncertainty of about 1\% in 
the absolute value of momentum as well as the using of different functional
forms to fit the points near the maxima.

\input epsf
\begin{figure}[ht]
  \begin{center}
    \mbox{\epsfxsize=7cm \epsfbox{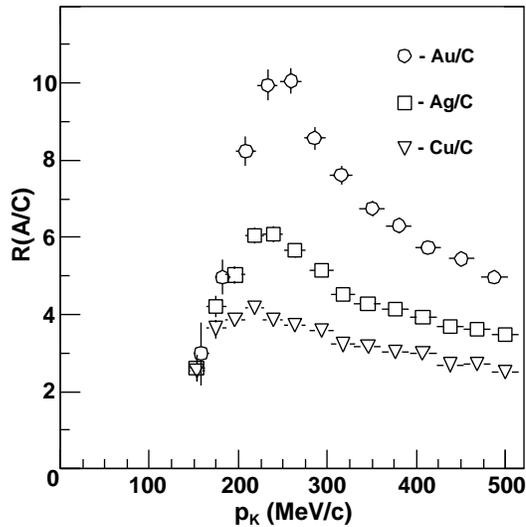}}
    \caption{Ratios of the $K^+$ production cross sections on Cu, Ag, and Au
      measured at $T_p=2.3\,$GeV as a function of the laboratory kaon
      momentum.}
  \end{center}
  \label{fig:2}
\end{figure}

It is clear from Fig.~1 that the suppression of $R(\AAu/\AC)$ at low
$p_K$ is largely independent of beam energy and of the angular 
acceptance of the spectrometer, suggesting that
the phenomenon is principally due to the interaction of the $K^+$ with
the residual nucleus. On the other hand, Fig.~2 shows that the
position of the maximum in $R(\AAA/\AC)$ increases with A. The situation
has a parallel in the well-known suppression of $\beta^+$ emission in 
heavy nuclei at low positron momenta arising from the repulsive Coulomb
field. Thus a $K^+$ produced at rest at some radius $R$ in the nucleus would,
in the absence of all other interactions, acquire a momentum of
$p_{\textrm{\scriptsize min}}=\sqrt{2m_KV_C(R)}$. Taking $R$ to be the
nuclear edge, this purely classical argument leads to a minimum $K^+$
momentum for Au of $p_{\textrm{\scriptsize min}}\approx 130\,$MeV/c.

It is, moreover, known from $K^+$ elastic scattering experiments at higher
energies~\cite{Marlow} that the $K^+A$ potential is mildly repulsive,
and this is in accord with one-body optical potentials based upon low-energy
$K^+N$ scattering parameters~\cite{Martin}. At normal nuclear density,
$\rho_0 \approx 0.16\,$fm$^{-3}$, the predicted repulsive $K^+A$ 
potential of strength $V_K^0\approx 20-25\,$MeV~\cite{Si97} would shift
$p_{\textrm{\scriptsize min}}$ to higher values.

In order to see whether the observed low momentum suppression is
compatible with such a combination of Coulomb and nuclear repulsion,
we performed calculations in the framework of the coupled channel 
transport model~\cite{Rudy,Ca99}. In this approach the different 
mechanisms for the kaon production and the influence of 
average Coulomb and nuclear potentials, as well as hadron rescattering 
effects, which can cause a sudden change of the kaon trajectory
when kaon comes close to a nucleon,
can be taken into account using realistic density distributions. Results 
of the calculations for the $R(\AAu/\AC)$ ratio are shown in
Fig.~3.  Without including the Coulomb and kaon potentials 
(dashed line in Fig.~3a) the ratio exhibits a smooth momentum
dependence with a steady increase towards low momenta resulting
from the stronger $K^+$ rescattering processes for the Au target. A
behaviour of this type was observed in $\pi^+$ production, which was
also measured in a short test experiment (see Fig.~1b). In this case
the influence of the Coulomb potential is expected to show up below
$p_{\pi}\approx 80\,$MeV/c, which was below our acceptance limit.  
For kaons the pure Coulomb interaction
leads to a distortion of the momentum spectrum and provides a maximum
at $p_K\approx 200\,$MeV/c (dashed-dotted line in Fig.~3a).
It has been checked that a 10\% change in the charge radius would 
move the maximum by less than $\pm 2\,$MeV/c.
 
A much larger change is observed in the calculations when a 
repulsive kaon nuclear 
potential is also considered, giving a 40 and 80~MeV/c shift with 
a potential strength of 20 and 40~MeV, respectively. When a kaon potential 
of $V_K^0= 20\,$MeV is used in the calculations, a reasonable
agreement with the experiment is achieved, with a maximum close to the 
experimental value of $245\pm 5\,$MeV/c (solid line in Fig.~3b). To 
take into account the absorption of the incident
proton, a baryon potential has also been included. 
The latter does not change the position of the maximum, but makes 
the agreement with the experimental data better.
A precision of 5~MeV/c in the position of the maximum, taken
together with the sensitivity of the data to kaon potential shown above,
might allow one to determine the strength of $V_K^0$ to better than 3~MeV.

\input epsf
\begin{figure}[ht]
  \begin{center}
   \mbox{\epsfxsize=7cm \epsfbox{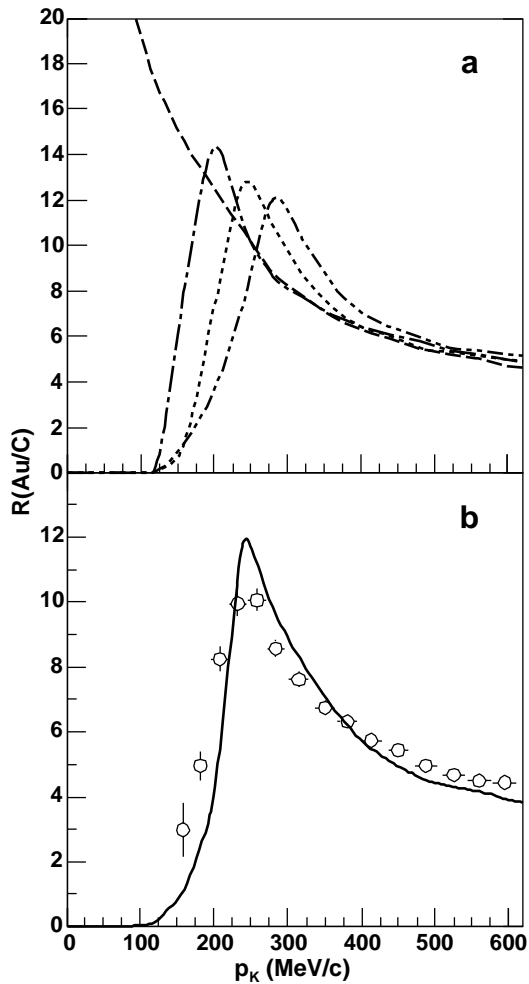}}
   \caption{Ratios of $K^+$ production cross sections for Au/C 
at $T_p=2.3\,\mathrm{GeV}$ as a function of the kaon momentum. 
a) The dash-dotted line is obtained from transport calculations including
only the Coulomb potential, the dotted line corresponds to 
calculations with the addition of a kaon potential of 20~MeV at $\rho_0$, 
whereas the dashed-double-dotted line shows the result where a kaon
potential of 40~MeV has been used. The broken line
corresponds to simulations without Coulomb and nuclear kaon potentials.
In all cases considered here, $K^+$ rescattering in the nucleus 
has been taken into account. b) The open circles are the experimental data.
The solid line shows the result of transport calculations starting from 
the dotted line in the top figure with a baryon potential added.}
  \end{center}
  \label{fig:cbuu}
\end{figure}

In summary, we have observed a strong suppression of the ratio of
$K^+$ production by protons on heavy nuclei to that on carbon at low
kaon momenta. The independence of this effect from beam energy and
the variation of the structure with $A$ provides clear evidence for
the influence of the $K^+$ Coulomb and nuclear interaction potentials.
The sensitivity found within our model suggests that a careful study 
of this region will provide a new way to investigate the $K^+A$ optical 
potential at low momenta. For this to be successful, more extensive 
transport calculations or other phenomenological descriptions have 
to be developed. Our preliminary analysis suggests that the $K^+$ 
nuclear potential at normal nuclear matter density is of the order 
of $20\,$MeV, which is in line with $K^+$ elastic scattering 
experiments~\cite{Marlow} and low-energy $K^+N$ scattering 
parameters~\cite{Martin}. Data on $K^+$ production from heavy-ion 
reactions at GSI also point towards $K^+$ nuclear potentials of 
about the same strength~\cite{Kaos,Skin}. We hope that the accuracy of
our data will stimulate further development of model calculations to
provide a better description of the momentum spectra.
If this is done we could expect the strength of the kaon potential to 
be extracted with an accuracy better than 3$\,$MeV.

We wish to acknowledge the assistance we received from the COSY staff
when performing these measurements at ANKE.  Financial support from
the following funding agencies was invaluable for our work: Georgia
(Department of Science and Technology), Germany (BMBF: grants
WTZ-POL-001-99, WTZ-RUS-649-96, WTZ-RUS-666-97, WTZ-RUS-685-99,
WTZ-POL-007-99; DFG: 436 RUS 113/337, 436 RUS 113/444, 436 RUS
113/561, State of North-Rhine Westfalia), Poland (Polish State
Committee for Scientific Research: 2 P03B 101 19), Russia (Russian
Ministry of Science, Russian Academy of Science: 99-02-04034,
99-02-18179a) and European Community (INTAS-98-500).

\end{document}